\documentclass[12pt]{article}
\usepackage{amsmath, amssymb}

\newcommand{\rh}{\rho}

\newcommand{\la}{\lambda}
\newcommand{\del}{\partial}
\newcommand{\de}{\delta}
\newcommand{\De}{\Delta}
\newcommand{\vphi}{\varphi}

\newcommand{\Ph}{M}
\newcommand{\ili}{\int\limits}
\newcommand{\nonu}{\nonumber}
\newcommand{\ww}{\nu}
\newcommand{\E}{{\rm e}}
\newcommand{\I}{{\rm i}}
\newcommand{\dd}{{\rm d}}
\newcommand{\ann}{a^{\hphantom{+}}}
\newcommand{\cre}{a^{+}}
\newcommand{\bnn}{b^{\hphantom{+}}}
\newcommand{\bre}{b^{+}}
\newcommand{\bZ}{{\mathbb Z}}
\newcommand{\bN}{{\mathbb N}}
\newcommand{\bC}{{\mathbb C}}
\newcommand{\bR}{{\mathbb R}}
\newcommand{\cA}{{\mathcal A}}
\newcommand{\cB}{{\mathcal B}}
\newcommand{\cF}{{\mathcal F}}
\newcommand{\cH}{{\mathcal H}}

\newcommand{\e}{\varepsilon}
\newcommand{\eng}{e}

\newcommand{\sfrac}[2]{{\textstyle \frac{#1}{#2}}}
\newcommand{\abs}[1]{{\left\vert #1 \right\vert}}
\newcommand{\Ref}[1]{$(\ref{#1})$}

\begin{document}
\title{
On the quantum 
Boltzmann equation
}

\author{L\'aszl\'o Erd\H os
\footnote{Work partially supported by NSF grant DMS-0200235}
\\ School of Mathematics, GeorgiaTech \\ \\
Manfred Salmhofer \\
Max-Planck Institute for Mathematics,\\
and Theoretical Physics, University of Leipzig \\ \\
Horng-Tzer Yau
\footnote{Work partially supported by NSF grant
DMS-0072098} \\
Courant Institute of Mathematical Sciences,  New York University\\
}

\maketitle

\begin{abstract}
 We give a nonrigorous derivation of the
nonlinear Boltzmann equation from the Schr\"odinger evolution of
interacting fermions. The argument is based mainly on the
assumption that a quasifree initial state satisfies a property
called {\it restricted quasifreeness} in the weak coupling  limit
at any later time.  By definition, a state is called restricted
quasifree if the four-point and the eight-point functions of the
state factorize  in the same manner as in a quasifree state.

\end{abstract}

\section{Introduction}
\noindent

The fundamental equation governing many-body quantum dynamics,
the Schr\"odinger equation, is a time reversible hyperbolic equation.
The quantum dynamics of many-body systems, however,  are often
modelled by a nonlinear, time irreversible (quantum)
Boltzmann equation, which  exhibits a particle-like
behavior. This apparent contradiction has
attracted a lot of attention over the years and, in particular,
Hugenholtz \cite{Hugenholtz} gave a derivation based
on a perturbation expansion
involving multiple commutators.
He selected a class of terms from this
expansion and argued that it gives the Boltzmann equation.
To the second order in the coupling constant,
Hugenholtz's claim was proved by Ho and Landau  \cite{Landau}.
Beyond that it is not even clear that these terms
satisfy the Boltzmann equation
order by order, partly due to the complicated selection rules.

In this paper, we present a derivation of the quantum Boltzmann
equation under the main assumption that in the weak coupling
limit \eqref{wl} the four-point and the eight-point functions of
the state factorize at any time  in the same manner as in a
quasifree state, see \eqref{4pf} and \eqref{8pf}. A state with
such a factorization property is called a {\it restricted
quasifree state}.
To rigorously verify this  assumption,
one has to analyze the connected $m$--point functions,
a very difficult problem in our view. So it might appear that
we have not improved much beyond the work \cite{Hugenholtz}.
Our approach however has the following two main merits. First:
It identifies the
concept, the restricted quasifreeness,
to  replace the independence in the classical setting
so that the structure of  the
collision term, i.e., the quartic nonlinearity and the product of
the factors $F$ and $1-F$ in the equation
\eqref{homB}, appears as a simple
consequence of this assumption.
Second: Unlike Hugenholtz's approach which is tied to
the commutator expansion, the restricted quasifreeness can now be verified
using  other  methods such as field--theoretical techniques.

Recent work by Benedetto, Castella, Esposito and Pulvirenti \cite{BCEP}
has given an interesting different derivation. We had learned this
work in a recent meeting and had subsequently sent them  an early version
of this manuscript.

%

\section{Definitions of the Dynamics}

We describe the quantum dynamics in the second--quantized
formulation. For definiteness, we shall restrict ourselves to a
fermion system. Our derivation is valid for bosons as well with
the only difference being that  some  $\pm$ signs change along the
derivation and in the final quantum Boltzmann equation (the terms
$(1-F)$ change to $(1+F)$ in \eqref{homB}). It should be noted
that many-boson systems are in general more difficult to control
rigorously than  many-fermion systems. The quantum Boltzmann
equation for fermions also preserves the property $0\leq F \leq
1$. On the contrary, the equation for bosons may blow up in finite
time.

Most of our setup is fairly standard and we recall the details
briefly here. For background, see
\cite{Bratteli,msren}.
The configuration space is a discrete torus $\Lambda=\bZ^d/L\bZ^d$,
of a very large sidelength, $L \in \bN$, which is kept finite throughout
the argument. The Hilbert space for the fermions is the standard
Fock space $\cF_\Lambda: = \bigoplus_{n \ge 0} \bigwedge^n \cH_\Lambda $,
where
$\cH_\Lambda : = \ell^2 (\Lambda, \bC)$.
\footnote{Thus the fermions are spinless.
We could also choose $\cH'_\Lambda  = \ell^2(\Lambda, \bC^2)$
as the one--particle Hilbert space,
to allow for a fermion spin $1/2$,
without changing the derivation in an essential way. }
Because $\cH_\Lambda $ is finite--dimensional,
the same holds for $\cF_\Lambda $.

We shall work in momentum space, which for our finite lattice
is the discrete torus
$\Lambda^*:= \sfrac{2\pi}{L} \bZ^d/2\pi \bZ^d$.
Let $w_p(x): = \E^{\I p \cdot x}$ where $p \cdot x := \sum_{i=1}^d p_i x_i$.
The set of functions $\{ w_p : p \in \Lambda^*\}$ is an orthogonal
basis of $\cH$ (the normalization is $\Vert w_p\Vert^2 = L^d$).
Therefore the annihilation operators $\ann_p = a(w_p)$ that are
associated to this basis in the standard way (see \cite{Bratteli}
or \cite{msren})
obey the canonical anticommutation relations (CAR)
\begin{equation}
\ann_p\cre_q + \cre_p\ann_q = \delta (p,q):
=
\begin{cases}
 L^d & \mbox{ if } p=q \\
0 & \mbox{ otherwise.}
\end{cases}
\end{equation}
If $F$ is continuous on $\cB:=\bR^d / 2\pi \bZ^d$, then
$L^{-d} \sum_{p \in \Lambda^*} F(p) \to
\int_\cB \sfrac{\dd^d p}{(2\pi)^d} \; F(p)$ as $L\to\infty$.
Since we are ultimately interested in the
limit $L \to \infty$,  we will use the continuum notation
even for the finite sums,
i.e.\ we write $\int_{\Lambda^*} \dd p\; F(p) $
for $L^{-d} \sum_{p \in \Lambda^*} F(p)$, etc.

Let $\cA$ be the $C^*$ algebra generated by
$\{\cre_p, \ann_p: p \in \Lambda^*\}$.
For a selfadjoint element $H \in \cA$,
consider the time evolution given by $H$ in the
Heisenberg picture, i.e.\ $A_t := \E^{-\I tH} A \E^{\I tH}$
for all $A \in \cA$.
Given a state $\rho$ on the algebra, we
define for $A \in \cA$, $\rh_t (A): = \rh(A_t)$.
For all $t \in \bR$, $\rh_t$ is again a state on $\cA$,
and its time evolution is given by the  Schr\"odinger equation
\begin{equation}
\I \sfrac{\del}{\del t} \rh_t (A) = \rh_t ( [H,A]).
\end{equation}
We take the Hamiltonian $H:=H_0 + \la \Phi$ where
\begin{equation}
H_0 := \int \dd p \; \eng(p) \, \cre_p \ann_p
\end{equation}
is the kinetic energy
and
\begin{equation}
\Phi:= \int \dd k_1 \ldots \dd k_4 \;
\langle k_1\; k_2 \mid \Phi \mid k_3\; k_4 \rangle
\cre_{k_1} \cre_{k_2} \ann_{k_3} \ann_{k_4}
\end{equation}
is the  interaction.
The coefficient function
$\langle k_1\; k_2 \mid \Phi \mid k_3\; k_4 \rangle$
is antisymmetric under exchange of $k_1$ and $k_2$
and under exchange of $k_3$ and $k_4$, and it contains
the momentum conservation delta function. For an
interaction generated by a two-body potential $v(x-y)$ we have
\begin{eqnarray} \label{Vanti}
\langle k_1\; k_2 \mid \Phi \mid k_3\; k_4   \rangle &=&
\de (k_1+k_2, k_3+k_4)
\qquad \qquad{}
\\
&& \frac14 \left( \hat v (k_1-k_4) - \hat v (k_2-k_4)
-\hat v (k_1-k_3) +\hat v (k_2-k_3) \right).
\nonu
\end{eqnarray}
We assume that
$\langle k_4\; k_3 \mid \Phi \mid k_2\; k_1    \rangle =
\overline{\langle k_1\; k_2 \mid \Phi \mid k_3\; k_4   \rangle}$.
Then $\Phi=\Phi^+$.
In terms of $v$ this condition 
means that $v$ is real. For simplicity,
we shall assume that $v$ is symmetric,
i.e.,
\begin{equation}\label{sym}
v(x) = v(-x) \ .
\end{equation}

 We shall call a polynomial $F$ in the creation and
annihilation operators {\it quartic} if it is homogeneous of
degree four and contains exactly two creation and two annihilation
operators. Any such quartic $F$ has a representation  similar to
\Ref{Vanti} with a coefficient $\langle k_1\; k_2 \mid F\mid k_3\;
k_4   \rangle $, and we shall always assume that the coefficient
is given in the properly antisymmetrized form so that we can
compare coefficients.

The state $\rh_t$ is determined by its values on monomials in the
creation and annihilation operators.
The two point function in Fourier space is defined by
\begin{equation}\label{2pt}
\ww_{pq}(t) := \rh_t(\cre_p \ann_q) .
\end{equation}
We are interested in the Euler scaling limit
of the two point function. In configuration space
it amounts to the rescaling
$$
x = X/\e,\; t = T/\e, \qquad \e\to 0.
$$
Recall the {\it Wigner transform} of a function $\psi\in L^2(\mathbb R^d)$
is
defined as
$$
        W_\psi(x,v): = \int e^{i\eta x}\overline{\widehat
\psi\Big(v-\frac{\eta}{2}\Big)}
        \widehat\psi\Big(v+\frac{\eta}{2}\Big) d\eta \;
         = \int e^{ivy} \overline{\psi\Big(x +\frac{y}{2}\Big)}
        \psi\Big( x-\frac{y}{2}\Big) dy \; .
$$
Define the rescaled Wigner distribution as
$$
        W^\e_\psi (X, V) : = \e^{-d}W_\psi\Big( \frac{X}{\e}, V\Big).
$$
Its  Fourier transform in $X$ is given by
$$
     \widehat W^\e_{\psi}(\xi, V) =
     \overline{\widehat \psi \Big(V-\frac{\e \xi}{2}\Big)}
        \widehat \psi \Big( V+ \frac{\e \xi}{2}\Big) \; .
$$
We can easily extend this notions to the two point function
$\ww_{pq}$. In particular, we can define the rescaled Wigner distribution
$W^\e_\rho(X, V)$ through its Fourier transform:
$$
     \widehat W^\e_\rho (\xi, V) = \rho(
     \cre_{V-\frac{\e \xi}{2}} \;  \ann_{V+ \frac{\e \xi}{2}} ) \; .
$$
Assume that
$$
W^\e(X, V, T):= W^\e_{\rho_{T/\e} }(X, V) \to F(X, V, T)
$$
as $\e \to 0$. Under the weak coupling scaling assumption, i.e.,
\begin{equation}\label{wl}
x = X/\e,\; t = T/\e, \; \lambda= \sqrt \e
\end{equation}
one expects that $F(X, V, T)$
satisfies the nonlinear Boltzmann equation
\begin{eqnarray}\label{inhomB}
\lefteqn{\frac{\del F(X, V, T)}{\del t} +  V \cdot\nabla_X F(X, V, T)}
\nonu\\
&=& 4\pi\int \dd k_2 \dd k_3\dd k_4 \; \de (k_1+k_2,k_3+k_4)\;
\de(E_1+E_2-E_3-E_4)
 \\
&&
\times\abs{\hat v(k_1-k_4) - \hat v (k_1-k_3)}^2 \nonu\\
&&\times\Big[F_{k_1}  F_{k_2 }  (1-F_{k_3 }) (1- F_{k_4 })- F_{k_4
} F_{k_3 }
 (1- F_{k_2 })(1- F_{k_1}) \Big] \nonu
\end{eqnarray}
where $F_{k_j}$ is short notation for $F(X, k_j,  T)$
and $E_i=\eng(k_i)$ with $k_1 = V$.
If  the state is homogeneous in space
(i.e.\ translation invariant)  at time zero,
then $\ww_{pq} = \de (p,q) F_p(t)$ for all later times $t$ as well
and the Boltzmann equation is reduced to
\begin{eqnarray}\label{homB}
 \frac{\del F_{k_1}}{\del T}
&=& 4\pi \int \dd k_2 \dd k_3\dd k_4 \; \abs{\hat v(k_1-k_4) - \hat v
(k_1-k_3)}^2
\\
&&
\times\de (k_1+k_2,k_3+k_4)\;
\de(E_1+E_2-E_3-E_4)
\nonu\\
&&
\times\lbrack
F_{k_1}  F_{k_2 }  (1-F_{k_3 }) (1- F_{k_4 })
-
F_{k_4 } F_{k_3 } (1- F_{k_2 })  (1- F_{k_1 })
\rbrack .
\nonu
\end{eqnarray}
Our goal is to give a heuristic derivation of this equation.

{\it Remark 1.} Without the symmetry assumption (\ref{sym}) one
has to replace the collision kernel $\abs{\hat v(k_1-k_4) - \hat
v(k_1-k_3)}^2$ in (\ref{homB}) with its symmetrized version
$$
     \frac{1}{4}  \abs{ \hat v (k_1-k_4) - \hat v (k_2-k_4)
-\hat v (k_1-k_3) +\hat v (k_2-k_3)}^2
$$
and our derivation remains valid.

The quartic structure of the collision is due to the quantum
nature and the weak coupling limit. Instead of the weak coupling
limit, one can take the low density limit ($ x = X/\e,\; t = T/\e,
\lambda=1$ and the density of the particles is $\e$). The
resulting equation will be the standard nonlinear Boltzmann
equation where collision term is quadratic with full quantum
scattering kernel and not just its Born approximation in the weak
coupling limit.
Technically, the emergence of the full quantum scattering kernel
can be seen from resumming the Born series, see \cite{EY} for the
simpler case of the Lorentz gas. From our experience working on
the weak coupling \cite {EY2} and low density limits in random
environments, we believe that a rigorous derivation of the low
density limit will be somewhat more complicated than in the weak
coupling limit. However,  the key difficulties arising from
many-body quantum dynamics are already present in the weak
coupling limit which  we shall focus on.

\bigskip
\section{The equation for the two point function}
\bigskip

The Schr\"odinger equation can be written as
\begin{equation}\label{Schnu}
(\I \del_t - \eng(p) + \eng(q)) \ww_{pq}(t) = \la \rh_t(F_{pq}),
\end{equation}
where $F_{pq}: = [\Phi, \cre_p \ann_q]$ is quartic with
\begin{eqnarray}
\langle k_1\; k_2 \mid F_{pq}\mid k_3\; k_4\rangle =
&-& \de(q,k_4) \langle k_1\; k_2 \mid \Phi \mid p\; k_3\rangle
\nonu\\
&+& \de(q,k_3) \langle k_1\; k_2 \mid \Phi \mid p\; k_4\rangle
\nonu\\
&+&\de(p,k_1) \langle k_2 \; q \mid \Phi \mid k_3\; k_4\rangle
\\
&-&\de(p,k_2) \langle k_1\; q \mid \Phi \mid k_3\; k_4 \rangle.
\nonu
\end{eqnarray}
Therefore 
\begin{equation}\label{Duh}
\ww_{pq}(t) = \ww_{pq}(0) \E^{-\I t(\eng(p)-\eng(q))}
- \I \la \ili_0^t \dd s \; \E^{-\I (t-s)(\eng(p)-\eng(q))}
\rh_s(F_{pq}) .
\end{equation}
Thus we need expectation values of quartic monomials, whose
evolution equation analogous to \Ref{Duh} involves
 ones of degree six, etc. This system of
equations ``hierarchy") is similar to the Schwinger--Dyson
equations, but the commutator structure implies that in an
expansion in Feynman graphs only connected graphs contribute.

Let $\vphi_{r_1,\ldots , r_4}(t ) :=
\rh_t( \cre_{r_1} \cre_{r_2} \ann_{r_3} \ann_{r_4})$ and
\begin{equation}
\De \eng (r_1, \ldots, r_4) := \eng(r_1) + \eng(r_2)
-\eng(r_3) -\eng(r_4).
\end{equation}
The Schr\"odinger equation for $\vphi$ gives
\begin{equation}
(\I \del_t - \De \eng (r_1, \ldots, r_4) )\vphi_{r_1,\ldots , r_4}(t
)
=
\la \rh_t( [\Phi, \cre_{r_1} \cre_{r_2} \ann_{r_3} \ann_{r_4}])
\end{equation}
which integrates to
\begin{eqnarray}
\vphi_{r_1,\ldots , r_4}(t ) &=&
\vphi_{r_1,\ldots , r_4}( 0)
\E^{-\I t  \De \eng (r_1, \ldots, r_4)}
\nonu
\\
&&- \I \la \ili_0^t \dd s \; \E^{-\I (t-s)  \De \eng (r_1, \ldots, r_4)}
\rh_s( [\Phi, \cre_{r_1} \cre_{r_2} \ann_{r_3} \ann_{r_4}]) .
\end{eqnarray}
Thus
\begin{equation}
\rh_t (F_{pq}) = \rh_0 (F_{pq}) - \I \la
\ili_0^t \dd s\; \rh_s ([\Phi,G_{pq}(t-s)])
\end{equation}
where $G_{pq} = G_{pq}(t-s)$ is quartic with
\begin{eqnarray}
\langle k_1\; k_2 \mid G_{pq} \mid k_3\; k_4\rangle
&=&
\E^{-\I (t-s)  \De \eng (k_1, \ldots, k_4)}
\\
&&
\Big[
\langle k_1\; k_2 \vert \Phi \vert k_3\; p\rangle \de(k_4,q)
- \langle k_1\; k_2 \vert \Phi \vert k_4\; p\rangle \de(k_3,q)
\nonu\\
&&-
\langle q\; k_2 \vert \Phi \vert k_3\; k_4\rangle \de(k_1,p)
+ \langle q\; k_1 \vert \Phi \vert k_3\; k_4\rangle \de(k_2,p)
\Big] .
\nonu
\end{eqnarray}
Thus the equation \Ref{Schnu} for $\ww_{pq}(t)$ can be written as
\begin{equation} \label{e1}
(\I \del_t - \eng(p) + \eng(q)) \ww_{pq}(t) =
\la \rh_0 (F_{pq}) - \I \la^2
\ili_0^t \dd s\; \rh_s ([\Phi,G_{pq}(t-s)]).
\end{equation}

\section{Restricted Quasifreeness}
\noindent
Up to this point, everything was exact; a heuristic derivation
of the Boltzmann equation now begins by treating the state $\rh_s$
as quasifree in \Ref{e1},
i.e.\ expressing the term on the right hand side
of \Ref{e1} as a product over $\ww_{pq}(s)$.

To this end, it is advantageous to avoid any contractions
in the commutator, i.e.\ simply leave it in the form
$[\Phi,G] = \Phi G - G \Phi$, which gives
\begin{eqnarray}\label{MG}
[\Phi,G_{pq}(t-s)] &=& \int \dd k_1\ldots \dd k_4 \dd l_1 \ldots \dd l_4
\cre_{k_1} \cre_{k_2}   \ann_{l_4} \ann_{l_3}
\cre_{k_3} \cre_{k_4}   \ann_{l_2} \ann_{l_1}
\nonu\\
&&
\Ph_{pq} (k_1,k_2,k_3,k_4,l_1,l_2,l_3,l_4)
\end{eqnarray}
with
\begin{eqnarray}  \label{VGpqcomm}
\Ph_{pq} (k_1,\ldots,l_4)  &=&
\Big[
\langle k_1\; k_2|\Phi|l_4\; l_3\rangle
\langle k_3\; k_4|G_{pq}(t-s)|l_2\; l_1\rangle
\\
&&-
\langle k_3\; k_4|\Phi|l_2\; l_1\rangle
\langle k_1\; k_2|G_{pq}(t-s)|l_4\; l_3\rangle
\Big] .
\nonu
\end{eqnarray}
We recall that expectation values of higher order monomials
in a quasifree state $\rh_s$ can be expressed by the two-point functions
(see the Appendix).
In particular, the four-point function is given by
the following determinant
\begin{equation}\label{4pf}
\rh_s(
\cre_{k_1}  \cre_{k_2}  \ann_{l_2} \ann_{l_1} )
=
 \left\vert
\begin{array}{rr}
\ww_{k_1l_1} & \ww_{k_1l_2}  \\
\ww_{k_2l_1} & \ww_{k_2l_2}  \\
\end{array}
\right\vert
\end{equation}
%
and the  eight-point function appearing in  \Ref{MG} is
\begin{equation}\label{8pf}
\rh_s(
\cre_{k_1} \cre_{k_2}   \ann_{l_4} \ann_{l_3}
\cre_{k_3} \cre_{k_4}   \ann_{l_2} \ann_{l_1} )
=
 \left\vert
\begin{array}{rrrr}
\ww_{k_1l_1} & \ww_{k_1l_2} & \ww_{k_1l_3} & \ww_{k_1l_4} \\
\ww_{k_2l_1} & \ww_{k_2l_2} & \ww_{k_2l_3} & \ww_{k_2l_4} \\
\ww_{k_3l_1} & \ww_{k_3l_2} & \tilde \ww_{k_3l_3} & \tilde \ww_{k_3l_4} \\
\ww_{k_4l_1} & \ww_{k_4l_2} & \tilde \ww_{k_4l_3} & \tilde \ww_{k_4l_4} \\
\end{array}
\right\vert
\end{equation}
Here each $ \ww_{kl} $ stands for  $ \ww_{kl}(s)$ and
$\tilde \ww_{kl} = - \de(k,l) + \ww_{kl} (s) $ appears in the  lower right
block because the monomial is not normal ordered.
{\it We shall call a state $\rh_s$ restricted quasifree if both
\eqref{4pf} and \eqref{8pf} are satisfied. We shall assume this
condition  in the limit $\lambda \to 0$}.

Return to the derivation of the Boltzmann equation.
A Laplace expansion of the determinant gives
\begin{eqnarray}
\rh_t(\cre_{k_1} \ldots \ann_{l_1})
&=&
\left\vert\begin{array}{rr}
\ww_{k_1l_1} & \ww_{k_1l_2}\\
\ww_{k_2l_1} & \ww_{k_2l_2}\\
\end{array}\right\vert
\;
\left\vert\begin{array}{rr}
\tilde \ww_{k_3l_3} & \tilde \ww_{k_3l_4} \\
\tilde \ww_{k_4l_3} & \tilde \ww_{k_4l_4} \\
\end{array}\right\vert
\nonu\\
&-&
\left\vert\begin{array}{rr}
\ww_{k_1l_1} & \ww_{k_1l_2}\\
\ww_{k_3l_1} & \ww_{k_3l_2}\\
\end{array}\right\vert
\;
\left\vert\begin{array}{rr}
\ww_{k_2l_3} & \ww_{k_2l_4} \\
\tilde \ww_{k_4l_3} & \tilde \ww_{k_4l_4} \\
\end{array}\right\vert
\nonu\\
&+&
\left\vert\begin{array}{rr}
\ww_{k_1l_1} & \ww_{k_1l_2}\\
\ww_{k_4l_1} & \ww_{k_4l_2}\\
\end{array}\right\vert
\;
\left\vert\begin{array}{rr}
\ww_{k_2l_3} & \ww_{k_2l_4} \\
\tilde \ww_{k_3l_3} & \tilde \ww_{k_3l_4} \\
\end{array}\right\vert
\nonu\\
&+&
\left\vert\begin{array}{rr}
\ww_{k_2l_1} & \ww_{k_2l_2}\\
\ww_{k_3l_1} & \ww_{k_3l_2}\\
\end{array}\right\vert
\;
\left\vert\begin{array}{rr}
\ww_{k_1l_3} & \ww_{k_1l_4} \\
\tilde \ww_{k_4l_3} & \tilde \ww_{k_4l_4} \\
\end{array}\right\vert
\\
&-&
\left\vert\begin{array}{rr}
\ww_{k_2l_1} & \ww_{k_2l_2}\\
\ww_{k_4l_1} & \ww_{k_4l_2}\\
\end{array}\right\vert
\;
\left\vert\begin{array}{rr}
\ww_{k_1l_3} & \ww_{k_1l_4} \\
\tilde \ww_{k_3l_3} & \tilde \ww_{k_3l_4} \\
\end{array}\right\vert
\nonu\\
&+&
\left\vert\begin{array}{rr}
\ww_{k_3l_1} & \ww_{k_3l_2}\\
\ww_{k_4l_1} & \ww_{k_4l_2}\\
\end{array}\right\vert
\;
\left\vert\begin{array}{rr}
\ww_{k_1l_3} & \ww_{k_1l_4} \\
\ww_{k_2l_3} & \ww_{k_2l_4} \\
\end{array}\right\vert .  \nonu
\end{eqnarray}
Noting that $\Ph_{pq} (k_1,\ldots,l_4)$ is antisymmetric under exchange of
$l_1$ and $l_2$ and under exchange of $l_3$ with $l_4$, and that the same
is
true for each of the six summands in the Laplace expansion, we see that we
may replace every $2\times2$ determinant by the product of the diagonal
elements
if we include a symmetry factor $4$. Moreover, $\Ph_{pq} (k_1,\ldots,l_4) $ is
antisymmetric
under $(k_1,k_2,l_4,l_3) \to (k_3,k_4,l_2,l_1)$ (this is just the
antisymmetry of the commutator in its two arguments), but the last of the
six summands, $\ww_{k_3l_1}\ww_{k_4l_2}\ww_{k_1l_3}\ww_{k_2l_4}$,
is symmetric, so it cancels out.
Graphically, this is the cancellation of the disconnected term.
Thus the term multiplying $\Ph_{pq} (k_1,\ldots,l_4)$ is
\begin{eqnarray}
4 &(& \ww_{k_1 l_1 } \ww_{k_2 l_2 }\tilde\ww_{k_3 l_3 }\tilde\ww_{k_4 l_4
}
\nonu\\
&-&
\ww_{k_1 l_1 } \ww_{k_3 l_2 }\ww_{k_2 l_3 }\tilde\ww_{k_4 l_4 }
\nonu\\
&+&
\ww_{k_1 l_1 } \ww_{k_4 l_2 }\ww_{k_2 l_3 }\tilde\ww_{k_3 l_4 }
\nonu\\
&+&
\ww_{k_2 l_1 } \ww_{k_3 l_2 }\ww_{k_1 l_3 }\tilde\ww_{k_4 l_4 }
\nonu\\
&-&
\ww_{k_2 l_1 } \ww_{k_4 l_2 }\ww_{k_1 l_3 }\tilde\ww_{k_3 l_4 } )
\\
= 4 &\{& \ww_{k_1 l_1 } \ww_{k_2 l_2 }\tilde\ww_{k_3 l_3 }\tilde\ww_{k_4
l_4 }
\nonu\\
&+&
(\ww_{k_3 l_2 }\tilde\ww_{k_4 l_4 }
- \ww_{k_4 l_2 } \tilde\ww_{k_3 l_4 } ) \;
( \ww_{k_1 l_3 }    \ww_{k_2 l_1 }  -
\ww_{k_1 l_1 } \ww_{k_2 l_3 } )  \}.
\nonu
\end{eqnarray}
Again, the last factor is antisymmetric with respect to an exchange of
$k_1$ and $k_2$ and an exchange of $k_3$ and $k_4$, so there is another
symmetry factor $4$, and
\begin{eqnarray}\label{44}
\varrho_s([\Phi,G_{pq}(t-s)])
 &=& \int \dd k_1\ldots \dd k_4 \dd l_1 \ldots \dd l_4
\Ph_{pq} (k_1,\ldots,l_4) \nonu
\\
&& 4 (\ww_{k_1 l_1 } \ww_{k_2 l_2 }\tilde\ww_{k_3 l_3 }\tilde\ww_{k_4 l_4
}
+ 4 \ww_{k_1 l_1 } \ww_{k_2 l_3 }\ww_{k_4 l_2 }\tilde\ww_{k_3 l_4 } ).
\end{eqnarray}

We remark that if we had used the commutator contraction to
express $[\Phi, G]$ in \Ref{MG}, then we would have needed to
evaluate only monomials of degree six on the state $\varrho_s$.
The calculation would have been longer because certain
cancellations would be
 less transparent. However this approach has the advantage that it
requires the quasifree factorization property of  $\varrho_s$
only for degree six monomials instead of degree eight.

\section{Spatial homogeneity}
\noindent
If we assume that the distribution is homogeneous in space
(i.e.\ translation invariant)  at time zero,
then $\ww_{pq} = \de (p,q) f_p(t)$ for all later times $t$ as well
by the translation invariance of $H$. In this case
 there are further simplifications: the term $\rh_0 (F_{pp})$ vanishes
and the $\eng (p) - \eng(q)$ term in the differential equation
also drops out. Moreover, for $p=q$
\begin{eqnarray}
\langle k_1\; k_2 \mid G_{pq} \mid k_3\; k_4\rangle
&= &
\E^{-\I (t-s)  \De \eng (k_1, \ldots, k_4)}
\langle k_1\; k_2 \vert \Phi \vert k_3\; k_4\rangle
\\
&&
\times(\de(p,k_4) + \de(p,k_3) - \de (p,k_2) - \de(p,k_1))
\nonu\end{eqnarray}
so we can compute the contribution of the first term
on the right hand side of \Ref{44} as
\begin{eqnarray}\label{first}
\lefteqn{
\int \dd k_1\ldots \dd l_4
\Ph_{pq} (k_1,\ldots,l_4)
4 \ww_{k_1 l_1 } \ww_{k_2 l_2 }\tilde\ww_{k_3 l_3 }\tilde\ww_{k_4 l_4}}
\qquad\qquad\qquad\qquad{}
\nonu\\
&=&
\int \dd k_1\ldots \dd k_4
\E^{-\I (t-s)  \De \eng (k_1, \ldots, k_4)}
\\
&&
\times 8 \vert \langle k_1\; k_2 \vert \Phi \vert k_3\; k_4\rangle \vert^2
\;
(\de (k_4,p) - \de (k_1,p))
\nonu\\
&&
\times (f_{k_1} f_{k_2 } \tilde f_{k_3 }\tilde f_{k_4 } -
f_{k_4 }f_{k_3 }\tilde f_{k_2 }\tilde f_{k_1 }) \nonu
\label{e22}
\end{eqnarray}
with $\tilde f_p = 1-f_p$.
The second term $16\Ph_{pq} (k_1,\ldots,l_4)
 \ww_{k_1 l_1 } \ww_{k_2 l_3 }\ww_{k_4 l_2 }\tilde\ww_{k_3 l_4 }$
in \Ref{44} drops out
 because with the assignment of momenta it is equal to
\begin{equation}
 32 (\de(k_1,p)-\de(k_3,p))\langle k_1 k_2 \vert \Phi \vert k_3 k_2 \rangle
\langle k_3 k_4 \vert \Phi \vert k_4 k_1 \rangle
\cos \lbrack(t-s) (\eng(k_3) - \eng(k_1))\rbrack ,
\nonu
\end{equation}
and this quantity
vanishes because the delta functions in both $\Phi$ factors are
$\de(k_1,k_3)$
and thus $\de(k_1,p)-\de(k_3,p)=0$.

Inserting \Ref{Vanti} into \Ref{first}, recalling
that in our finite volume, $\de (p,p) = L^d$, so that
$\de(p,q)^2= L^d \de (p,q)$, and using (\ref{sym})
and  symmetry arguments as above, we get
\begin{eqnarray}\label{fp}
\del_t f_p(t) &=& - \la^2 \ili_0^t \dd s \int \dd k_1 \ldots \dd k_4 \;
\de (k_1+k_2,k_3+k_4)
\E^{-\I (t-s)  \De \eng (k_1, \ldots, k_4)}
\nonu\\
&&
\times 2 (\de (k_4,p) - \de (k_1,p))
\abs{\hat v(k_1-k_4) - \hat v (k_1-k_3)}^2
\\
&&
\times(f_{k_1}(s) f_{k_2 }(s) \tilde f_{k_3 }(s)\tilde f_{k_4 }(s) -
f_{k_4 }(s)f_{k_3 }(s)\tilde f_{k_2 }(s)\tilde f_{k_1 }(s)) .
\nonu
\end{eqnarray}

\section{Local approximation in time}
\noindent
We rewrite the equation as
\begin{equation}\label{28}
-\la^{-2} \del_t f_p(t) = \ili_{-\infty}^{\infty} \dd E
\ili_0^t \dd s \; \E^{-\I E (t-s)} \; \beta(E,p,s)
\end{equation}
with
\begin{eqnarray}\label{B}
\beta(E,p,s) &=&
\int \dd k_1 \ldots \dd k_4 \;
\de (k_1+k_2,k_3+k_4)\; 2 (\de (k_4,p) - \de (k_1,p))
\nonu\\
&&
\times\abs{\hat v(k_1-k_4) - \hat v (k_1-k_3)}^2
\de(E-\De \eng(k_1,\ldots,k_4))
\\
&&
\times (f_{k_1}(s) f_{k_2 }(s) \tilde f_{k_3 }(s)\tilde f_{k_4 }(s) -
f_{k_4 }(s)f_{k_3 }(s)\tilde f_{k_2 }(s)\tilde f_{k_1 }(s)) ,
\nonu
\end{eqnarray}
Since $v$ is symmetric \eqref{sym}, $\beta$ is a symmetric
function of $E$.

Notice that $f$ and $\beta$ are $\lambda$ dependent and we shall
denote them by $f^\lambda$ and $\beta^\lambda$. We now assume that
the limits
\begin{equation}\label{limit}
\lim_{\lambda \to 0} f_p^\lambda(T/\lambda^2)= F(T, p), \qquad
\lim_{\lambda \to 0} \beta^\lambda(E, p, T/\lambda^2)= B(E,  p, T)
\end{equation}
exist and the relation \eqref{B} continues to hold in the limit.
We can take the limit  $\lambda \to 0$ in  \eqref{28} and this yields
\begin{equation}\label{30}
-\del_T F(T, p) = \lim_{\lambda \to 0} \ili_{-\infty}^{\infty} \dd E
\ili_0^{T}   \frac {\dd S} {\lambda^2}  \;
 \E^{-\I E (T-S)/\lambda^2} \; \beta^\lambda(E,p,S/\lambda^2) \ .
\end{equation}
We now assume that we can replace the function
$\beta^\lambda$ by its limit $B$. Thus we have
\begin{equation}\label{31}
-\del_T F(T, p) = \lim_{\lambda \to 0} \ili_{-\infty}^{\infty} \dd E
\ili_0^{T}   \frac {\dd S} {\lambda^2}  \;
 \E^{-\I E (T-S)/\lambda^2} \; B (E,p,S/\lambda^2)\ .
\end{equation}
Interchanging the the integration and performing the $E$ integration, we have
\begin{equation*}
-\del_T F(T, p) = \lim_{\lambda \to 0}
\ili_0^{T}  \frac {\dd S} {\lambda^2}  \;
 \; \widehat B \big( \, (T-S)/\lambda^2, p,S/\lambda^2 \, \big )\ .
\end{equation*}
Let $u= (T-S)/\lambda^2$. We can rewrite the last equation as
\begin{equation*}
-\del_T F(T, p) = \lim_{\lambda \to 0}
\ili_0^{T}  \dd u  \;
 \; \widehat B \big( \, u, p, T+ u \lambda^2 \, \big ) \ .
\end{equation*}
In the limit $\lambda \to 0$, the right side converges to
$$
\ili_0^{\infty}  \dd u  \;
 \; \widehat B \big( \, u, p, T \, \big ) =
\frac{1}{2}\ili_{-\infty}^{\infty}  \dd u  \;
 \; \widehat B \big( \, u, p, T \, \big )
= \pi B(0, p, T) \ .
$$
where we have used the symmetry of $\beta$ in $E$.  Combining the
last two equations, we have derived the Boltzmann equation
\eqref{homB}.

In this derivation, we used
the restricted quasifreeness assumption, spatial homogeneity and the existence
of the limit for the the two point function $\nu_{pq}$ ({\textit{cf}}:
\eqref{limit}). We have not made precise the meaning of the limit
and we have freely interchanged limits with differentiations
and integrations etc. This suggests
that for a rigorous proof
the two point function has to be controlled precisely, perhaps
through some expansion method.

The spatial homogeneity of the initial state can be replaced
with the assumption that the point function at any time
scales as
$$
     \nu_{pq}(t) = R(\e t, \sfrac{p+q}{2}, \sfrac{p-q}{2\e}) .
$$
Our derivation above can easily be extended to this case to give
the spatially inhomogeneous Boltzmann equation \Ref{inhomB}.

\appendix
\section{Quasifree states and determinants}
\noindent
For finite $L$, the observable algebra is finite--dimensional, so a state $\rho$ is quasifree
if and only if it is given by density matrix coming from a quadratic Hamiltonian
(see, e.g.\ \cite{BLS}). That is,
\begin{equation}
\rho (A) = \frac1Z \mbox{ tr } (\E^{-H_0} A ) \ ,
\end{equation}
where $Z = $ tr $\E^{-H_0}$.
We restrict to states which are invariant under the transformations
$ \ann_{p} \to \E^{\I \alpha} \ann_p$ for all $\alpha \in \bR$.
For this case we prove below that the expectation value of any
normal ordered monomial can be computed with the following formula:
\begin{equation}\label{detrho}
\rho \left(
\prod\limits_{n=1}^m \cre_{p_n} \;
\prod\limits_{n'=1}^{m'} \ann_{q_{n'}}
\right)
=\de_{mm'}
(-1)^{m(m-1)/2}
\det \left(
\rho (\cre_{p_n} \ann_{q_{n'}})
\right)_{1 \le n,n'\le m} .
\end{equation}
To simplify notation, we enumerate our finite set $\Lambda^*$ in some way
so that we can replace the subscript $p \in \Lambda^*$ by a number
$i \in \{1, \ldots, N \}$, $N=L^d$. Moreover, because (\ref{detrho})
 is homogeneous,
we may rescale the creation and annihilation operators by $L^{-d/2}$,
so that they obey the CAR $\ann_i \cre_j + \cre_j \ann_i = \delta_{ij}$
with $\delta_{ij}$ the Kronecker delta.
With these conventions, and by the just stated $U(1)$ invariance,
\begin{equation}
H_0 = \sum_{i,j} \cre_i Q_{ij} \ann_j .
\end{equation}
Positivity of $\rho $ requires $H_0$ to be hermitian, so
$\bar Q_{ij} = Q_{ji}$. Thus there is $U \in U(N)$ such that
$Q= U E U^*$ with $E = \mbox{ diag }\{ E_1,\ldots, E_N\}$.
The operators $\bnn_k = \sum_j U_{kj} a_j$ have canonical anticommutation
relations $\bnn_k \bre_l + \bre_l \bnn_k = \delta_{kl}$,
so that $n_k = \bre_k \bnn_k$ satisfies $n_k^2 = n_k$ and
$n_k n_l = n_l n_k$. Thus $\E^{-H_0}$ is the product of commuting factors
\begin{equation}
\E^{-H_0} = \prod_k \left( 1+ (\E^{-E_k} -1) n_k \right),
\end{equation}
hence $Z= \prod_k (1+ \E^{-E_k})$, and $\rho (\bre_k \bnn_l) = \delta_{kl}
(1+\E^{E_k})^{-1}$. This implies
\begin{equation}
\rho(\cre_i \ann_{i'}) =
 U_{i'l} \frac{1}{1+\E^{E_l}}\bar U_{il}
= {(1+ \E^{Q})^{-1}}_{i'i}.
\end{equation}
Because $H_0$ is diagonal when expressed in terms of $\bre$ and $\bnn$,
\begin{equation}\label{perm}
\rho \left(
\prod\limits_{k=1}^m \bre_{u_k} \prod\limits_{k'=1}^{m'} \bnn_{v_k'}
\right)
\end{equation}
vanishes unless $m'=m$ and $(v_1, \ldots, v_m)$ is a permutation of
$(u_1, \ldots, u_m): \;$ $v_k=u_{\pi(k)}$. In that case, by the CAR,
(\ref{perm}) equals
\begin{equation}
(-1)^{m(m-1)/2} \mbox{ sign }(\pi) \;
\rho \left(
\prod\limits_{k=1}^m (\bre_{u_k} \bnn_{u_k})
\right) \ .
\end{equation}
Eq.\ (\ref{detrho}) now follows straightforwardly
by expressing the product of $\cre$ and $\ann$
in terms of $\bre$ and $\bnn$ and using the definition of the determinant.

In (\ref{4pf}) and (\ref{8pf}), the indices of the annihilation operators are ordered
downwards, so the factor $(-1)^{m(m-1)/2}$ is absent.
The procedure of commuting a monomial that is not normal ordered
to its normal ordered form corresponds to successive row expansions
of the determinant in (\ref{8pf}).

\noindent Addresses of the authors:

\medskip

\noindent  L\'aszl\'o Erd\H os \\
School of Mathematics\\
 GeorgiaTech Atlanta, GA 30332, U.S.A.\\
 lerdos@math.gatech.edu

\bigskip

\noindent Manfred Salmhofer \\
Max-Planck Institute for Mathematics, \\
 Inselstr. 22, D-04103 Leipzig, Germany\\
and \\
Theoretical Physics, University of Leipzig, \\
 Augustusplatz 10, D-04109 Leipzig, Germany\\
mns@mis.mpg.de

\bigskip

\noindent Horng-Tzer Yau\\
Courant Institute of Mathematical Sciences, \\
 New York University,\\
 New York, NY, 10012, U.S.A.\\
 yau@cims.nyu.edu

\end{document}